\documentstyle[prl,aps,multicol]{revtex}
\renewcommand{\narrowtext}{\begin{multicols}{2} \global\columnwidth20.5pc}
\renewcommand{\widetext}{\end{multicols} \global\columnwidth42.5pc}
\begin{document}
\draft
\title{Two--band random matrices}
\author{E. Kanzieper\thanks{%
Present address: Condensed Matter Section, The Abdus Salam International
Centre for Theoretical Physics, P.O. Box 586, 34100 Trieste, Italy } and V.
Freilikher}
\address{The Jack and Pearl Resnick Institute of Advanced Technology,\\
Department of Physics, Bar--Ilan University, 52900 Ramat--Gan, Israel}
\date{December 22, 1997}
\maketitle

\begin{abstract}
Spectral correlations in unitary invariant, non--Gaussian ensembles of large
random matrices possessing an eigenvalue gap are studied within the
framework of the orthogonal polynomial technique. Both local and global
characteristics of spectra are directly reconstructed from the recurrence
equation for orthogonal polynomials associated with a given random matrix
ensemble. It is established that an eigenvalue gap does not affect the local
eigenvalue correlations which follow the universal sine and the universal
multicritical laws in the bulk and soft--edge scaling limits, respectively.
By contrast, global smoothed eigenvalue correlations do reflect the presence
of a gap, and are shown to satisfy a new universal law exhibiting a sharp
dependence on the odd/even dimension of random matrices whose spectra are
bounded. In the case of unbounded spectrum, the corresponding universal
`density--density' correlator is conjectured to be generic for chaotic
systems with a forbidden gap and broken time reversal symmetry.
\end{abstract}

\pacs{\tt cond-mat/9709309}

\narrowtext

\section{Introduction}

Ensembles of large random matrices ${\bf H}$ generated by the joint
distribution function $P\left[ {\bf H}\right] \propto \exp \left\{ -\beta 
\mathop{\rm Tr}
V\left[ {\bf H}\right] \right\} $, with $\beta $ being a symmetry parameter
as explained below, may display phase transitions under non--monotonic
deformation of the confinement potential $V\left[ {\bf H}\right] $.
Different phases are characterized by topologically different arrangements
of eigenvalues in random matrix spectra that may have multiple--band
structure. Random matrices, whose spectra undergo phase transitions, appear
in quantizing two--dimensional gravity \cite
{Molinari1988,DSS-1990,Sasaki1991}, in the context of quantum chromodynamics 
\cite{Jurk-1996,JV-1996}, as well as in some models of particles interacting
in high dimensions \cite{Cugliandolo1995}. Transition regimes realized in
invariant random matrix ensembles have implications for a certain class of
Calogero--Sutherland--Moser models \cite{Morita1995}. These matrix models
may also be applicable to chaotic systems having a forbidden gap in the
energy spectrum.

In the eigenvalue representation, the invariant random matrix model is
defined by the joint probability distribution function \cite{Mehta-1991} 
\begin{equation}
P\left( \left\{ \varepsilon \right\} \right) ={\cal Z}_N^{-1}\prod_{i>j=1}^N%
\left| \varepsilon _i-\varepsilon _j\right| ^\beta \prod_{k=1}^N\exp \left\{
-\beta V\left( \varepsilon _k\right) \right\}  \label{q.01}
\end{equation}
of $N$ eigenvalues $\left\{ \varepsilon \right\} =\left\{ \varepsilon
_1,...,\varepsilon _N\right\} $ of an $N\times N$ random matrix ${\bf H}$.
The symmetry parameter $\beta $ coincides with a number of independent
elements in off--diagonal entries of a random matrix ${\bf H}$. For real
symmetric matrices, $\beta =1$ (orthogonal symmetry), $\beta =2$ for
Hermitian matrices (unitary symmetry), and $\beta =4$ for self--dual
Hermitian matrices (symplectic symmetry). It is convenient to parametrize
the confinement potential $V\left( \varepsilon \right) $ entering Eq. (\ref
{q.01}) by a set of coupling constants $\left\{ d\right\} =\left\{
d_1,...,d_p\right\} $, 
\begin{equation}
V\left( \varepsilon \right) =\sum_{k=1}^p\frac{d_k}{2k}\varepsilon
^{2k},\;d_p>0,  \label{q.02}
\end{equation}
so that we may consider the phase transitions as occurring in $\left\{
d\right\} $--space. Because the confinement potential is an even function,
the associated random matrix model possesses so--called $Z2$--symmetry.

Variations of the coupling constants affect the Dyson density $\nu _D$, that
can be found by minimizing the free energy ${\cal F}_N=-\log {\cal Z}_N$,
Eq. (\ref{q.01}), subject to a normalization constraint $\int \nu _D\left(
\varepsilon \right) d\varepsilon =N$, 
\begin{equation}
\frac{dV}{d\varepsilon }-{\cal P}\int d\zeta \frac{\nu _D\left( \zeta
\right) }{\varepsilon -\zeta }=0,  \label{q.03}
\end{equation}
where ${\cal P}$ indicates a principal value of the integral. When all $d_k$
are positive, so that confinement potential is monotonic, the spectral
density $\nu _D$ has a single--band support, ${\cal N}_{\text{b}}=1$.
Non--monotonic deformation of the confinement potential can be carried out
by changing the signs of some of $d_k$ $\left( k\neq p\right) $. Such a {\it %
continuous} variation of coupling constants may lead, under certain
conditions, to a {\it discontinuous} change of the topological structure of
spectral density $\nu _D$, when the eigenvalues $\left\{ \varepsilon
\right\} $ are arranged in ${\cal N}_{\text{b}}>1$ `allowed' bands separated
by `forbidden' gaps.

The phase structure of Hermitian $\left( \beta =2\right) $ one--matrix model
Eq. (\ref{q.01}) has been studied in a number of works \cite
{Shim,Demeterfi-1990,Cicuta-1990,Jurk-1991}, where the simplest examples of
non--monotonic quartic and sextic confinement potentials have been examined.
It has been found that there are domains in the phase space of coupling
constants where only a particular solution for $\nu _D$ exists, and it has a
fixed number ${\cal N}_{\text{b}}$ of allowed bands. However, in some
regions of the phase space, one can have more than one kind of solution of
the saddle-point equation Eq. (\ref{q.03}). In this situation, solutions
with different number of bands ${\cal N}_{\text{b}}^{\left( 1\right) },$ $%
{\cal N}_{\text{b}}^{\left( 2\right) },...$ are present simultaneously. When
such an overlap appears, one of the solutions, say ${\cal N}_{\text{b}%
}^{\left( k\right) }$, has the lowest free energy ${\cal F}_N^{\left(
k\right) }$, and this solution is dominant, while the others are
subdominant. Moreover, numerical calculations \cite{Jurk-1991} showed that
some special regimes exist in which the {\it bulk} spectral density obtained
as a solution to the saddle--point equation Eq. (\ref{q.03}) differs
significantly from the genuine level density computed numerically within the
framework of the orthogonal polynomial technique. It was then argued that
such a genuine density of levels cannot be interpreted as a multi--band
solution with an integer number of bands. A full understanding of this
phenomenon is still absent.

Recently, interest was renewed in multi--band regimes in invariant random
matrix ensembles. An analysis based on a loop equation technique \cite
{AmbjornAkemann1996,Akemann1996} showed that fingerprints of phase
transitions appear not only in the Dyson density but also in the (universal)
wide--range eigenvalue correlators, which in the multi--band phases differ
from those known in the single--band phase \cite
{AmbjornJurkMak1990,BrezinZee1993,Beenakker1994}. A renormalization group
approach developed in Ref. \cite{Higuchi1996} supported the results found in
Refs. \cite{AmbjornAkemann1996,Akemann1996} for the particular case of two
allowed bands, referring a new type of universal wide--range eigenlevel
correlators to an additional attractive fixed point of a renormalization
group transformation.

The method of loop equations \cite{AmbjornAkemann1996,Akemann1996}, used for
a treatment of non--Gaussian, unitary invariant, random matrix ensembles
fallen in a multi--band phase, is only suitable for computing the global
characteristics of spectrum. Therefore, an appropriate approach is needed
capable of analyzing local characteristics of spectrum (manifested on the
scale of a few eigenlevels). A possibility to probe the local properties of
eigenspectrum is offered by the method of orthogonal polynomials. A step in
this direction was taken in a recent paper \cite{Deo-1997}, where an ansatz
was proposed for large--$N$ asymptotes of orthogonal polynomials associated
with a random matrix ensemble having two allowed bands in its spectrum.
Because the asymptotic formula proposed there is of the Plancherel--Rotach
type \cite{Szego}, it is only applicable for studying eigenvalue
correlations in the spectrum bulk and cannot be used for studying local
correlations in an arbitrary spectrum range (for example, near the edges of
two--band eigenvalue support).

The aim of the present paper is to develop a new approach (within an
orthogonal polynomial scheme) allowing a unified treatment of eigenlevel
correlations in the unitary invariant $%
\mathop{\rm U}
\left( N\right) $ matrix model $\left( \beta =2\right) $ with a forbidden
gap. This is a further extension of the Shohat method \cite
{Shohat,Bonan-Clark} that has been used previously by the authors to study $%
\mathop{\rm U}
\left( N\right) $ invariant ensembles of large random matrices in the
single--band phase \cite{KF-philmag,KF-1997}. In particular, we are able to
study both the fine structure of local characteristics of the spectrum in
different scaling limits and smoothed global spectral correlations. Our
treatment is based on the direct reconstruction of spectral correlations
from the recurrence equation for the corresponding orthogonal polynomials.

\section{General relations}

In this section we briefly review the orthogonal polynomial technique \cite
{Mehta-1991}. The $n$--point correlation function that describes the
probability density to find $n$ levels around each of the points $%
\varepsilon _1,...,\varepsilon _n$ when the positions of the remaining
levels are unobserved is defined by the formula 
\begin{equation}
R_n\left( \varepsilon _1,...,\varepsilon _n\right) =\frac{N!}{\left(
N-n\right) !}\int_{-\infty }^{+\infty }P\left( \left\{ \varepsilon \right\}
\right) \prod_{k=n+1}^Nd\varepsilon _k.  \label{q.04}
\end{equation}
This correlation function can explicitly be expressed in terms of the
two--point kernel $K_N\left( \varepsilon ,\varepsilon ^{\prime }\right) $ as
follows 
\begin{equation}
R_n\left( \varepsilon _1,...,\varepsilon _n\right) =\det \left\| K_N\left(
\varepsilon _i,\varepsilon _j\right) \right\| _{i,j=1...n}\text{.}
\label{q.05}
\end{equation}
Here, 
\begin{equation}
K_N\left( \varepsilon ,\varepsilon ^{\prime }\right) =c_N\frac{\varphi
_N\left( \varepsilon ^{\prime }\right) \varphi _{N-1}\left( \varepsilon
\right) -\varphi _N\left( \varepsilon \right) \varphi _{N-1}\left(
\varepsilon ^{\prime }\right) }{\varepsilon ^{\prime }-\varepsilon }\text{,}
\label{q.06}
\end{equation}
and the `eigenfunctions' 
\begin{equation}
\varphi _n\left( \varepsilon \right) =P_n\left( \varepsilon \right) \exp
\left\{ -V\left( \varepsilon \right) \right\}  \label{q.07}
\end{equation}
are determined by the set of polynomials orthogonal with respect to the
measure $d\mu \left( \varepsilon \right) =\exp \left\{ -2V\left( \varepsilon
\right) \right\} d\varepsilon $, 
\begin{equation}
\int_{-\infty }^{+\infty }d\mu \left( \varepsilon \right) P_n\left(
\varepsilon \right) P_m\left( \varepsilon \right) =\delta _{nm},
\label{q.08}
\end{equation}
and obeying the recurrence equation 
\begin{equation}
\varepsilon P_{n-1}\left( \varepsilon \right) =c_nP_n\left( \varepsilon
\right) +c_{n-1}P_{n-2}\left( \varepsilon \right) .  \label{q.09}
\end{equation}
The recurrence coefficients $c_n$ entering Eqs. (\ref{q.06}) and (\ref{q.09}%
) are uniquely determined by the measure $d\mu $. Equations (\ref{q.05}) and
(\ref{q.06}) demonstrate that the problem of eigenvalue correlations is
reduced to that of finding asymptotes for the eigenfunctions $\varphi _N$.

\section{Mapping recurrence equation onto differential equation}

To map a recurrence Eq. (\ref{q.09}) onto a second--order differential
equation for eigenfunctions $\varphi _n$, we note that the first derivative $%
dP_n/d\varepsilon $ can be represented as \cite{Shohat,Bonan-Clark} 
\begin{equation}
\frac{dP_n}{d\varepsilon }=A_n\left( \varepsilon \right) P_{n-1}-B_n\left(
\varepsilon \right) P_n,  \label{q.10}
\end{equation}
where 
\begin{equation}
A_n\left( \varepsilon \right) =2c_n\int d\mu \left( t\right) \frac{V^{\prime
}\left( t\right) -V^{\prime }\left( \varepsilon \right) }{t-\varepsilon }%
P_n^2\left( t\right) ,  \label{q.11}
\end{equation}
\begin{equation}
B_n\left( \varepsilon \right) =2c_n\int d\mu \left( t\right) \frac{V^{\prime
}\left( t\right) -V^{\prime }\left( \varepsilon \right) }{t-\varepsilon }%
P_n\left( t\right) P_{n-1}\left( t\right) .  \label{q.12}
\end{equation}
Then, by using Eqs. (\ref{q.09}) and (\ref{q.10}), one obtains after some
algebra that the fictitious wave function $\varphi _n$ given by Eq. (\ref
{q.07}) satisfies the following differential equation: 
\begin{equation}
\frac{d^2\varphi _n\left( \varepsilon \right) }{d\varepsilon ^2}-{\cal F}%
_n\left( \varepsilon \right) \frac{d\varphi _n\left( \varepsilon \right) }{%
d\varepsilon }+{\cal G}_n\left( \varepsilon \right) \varphi _n\left(
\varepsilon \right) =0.  \label{q.13}
\end{equation}
Here, 
\begin{equation}
{\cal F}_n\left( \varepsilon \right) =\frac 1{A_n}\frac{dA_n}{d\varepsilon },
\label{q.14}
\end{equation}
and 
\begin{eqnarray}
{\cal G}_n\left( \varepsilon \right) &=&\frac{dB_n}{d\varepsilon }+\frac{c_n%
}{c_{n-1}}A_nA_{n-1}  \nonumber \\
&&\ -B_n\left( B_n+2\frac{dV}{d\varepsilon }+\frac 1{A_n}\frac{dA_n}{%
d\varepsilon }\right)  \nonumber \\
&&\ +\frac{d^2V}{d\varepsilon ^2}-\left( \frac{dV}{d\varepsilon }\right) ^2-%
\frac 1{A_n}\frac{dA_n}{d\varepsilon }\frac{dV}{d\varepsilon }.  \label{q.15}
\end{eqnarray}

Equation (\ref{q.13}) is valid for arbitrary $n$. We note that despite the
generality of the differential equation obtained, its practical use is quite
restricted since the functions ${\cal F}_n\left( \lambda \right) $ and $%
{\cal G}_n\left( \lambda \right) $ entering Eqs. (\ref{q.13}) can be
calculated explicitly only for rather simple measures $d\mu $. Nevertheless,
an asymptotic analysis of this equation is available in the limit $n=N\gg 1$%
, that is of great interest in random matrix theory.

\subsection{Single--band phase}

The single--band phase corresponds to monotonic confinement potentials or to
those having light local extrema. Corresponding asymptotic analysis has been
carried out by the authors in Refs. \cite{KF-philmag,KF-1997}. For further
comparison with a two--band--phase solution, we give a differential equation
for $\varphi _N^{\left( \text{I}\right) }\left( \varepsilon \right) $
obtained in the leading order in $N\gg 1$ [upper index indicates that the
single--band phase is considered]: 
\[
\frac{d^2\varphi _N^{\left( \text{I}\right) }\left( \varepsilon \right) }{%
d\varepsilon ^2}-\left[ \frac d{d\varepsilon }\log \left( \frac{\pi \nu
_D^{\left( \text{I}\right) }\left( \varepsilon \right) }{\sqrt{{\cal D}%
_N^2-\varepsilon ^2}}\right) \right] \frac{d\varphi _N^{\left( \text{I}%
\right) }\left( \varepsilon \right) }{d\varepsilon } 
\]
\begin{equation}
+\left[ \pi \nu _D^{\left( \text{I}\right) }\left( \varepsilon \right)
\right] ^2\varphi _N^{\left( \text{I}\right) }\left( \varepsilon \right) =0.
\label{q.16}
\end{equation}
It is remarkable that Eq. (\ref{q.16}) does not contain the confinement
potential explicitly, but only involves the Dyson density 
\begin{equation}
\nu _D^{\left( \text{I}\right) }\left( \varepsilon \right) =\frac 2{\pi ^2}%
{\cal P}\int_0^{{\cal D}_N}\frac{tdt}{t^2-\varepsilon ^2}\frac{dV}{dt}\sqrt{%
\frac{1-\varepsilon ^2/{\cal D}_N^2}{1-t^2/{\cal D}_N^2}}  \label{q.17}
\end{equation}
corresponding to the single--band phase and analytically continued on the
entire real axis; ${\cal D}_N$ is the soft edge of the spectrum, being the
positive root of the integral equation 
\begin{equation}
\int_0^{{\cal D}_N}\frac{dV}{dt}\frac{tdt}{\sqrt{{\cal D}_N^2-t^2}}=\frac{%
\pi N}2\text{.}  \label{q.18}
\end{equation}
It has been shown that for non--singular confinement potential, solutions of
Eq. (\ref{q.16}) lead to the universal sine kernel in the bulk scaling
limit, and to the so-called $%
\mathop{\rm G}
$--multicritical correlations in the soft--edge scaling limit \cite{KF-1997}%
. An additional logarithmic singularity of confinement potential introduces
additional terms into Eq. (\ref{q.16}), giving rise to the universal Bessel
correlations in the origin scaling limit \cite{NBgroup1997,KF-philmag}. For
further progress in the field, see very recent paper \cite{ADMN}. \widetext

\subsection{Two--band phase}

Let us consider the situation when the confinement potential has two deep
wells leading to the Dyson density supported on two disjoint intervals
located symmetrically about the origin, ${\cal D}_N^{-}<\left| \varepsilon
\right| <{\cal D}_N^{+}$. In this situation, the recurrence coefficients $%
c_n $ entering Eq. (\ref{q.09}) are known to be double--valued functions of
the number $n$ \cite{Molinari1988,Demeterfi-1990}. This means that for $%
n=N\gg 1$, one must distinguish between coefficients $c_{N\pm 2q}\approx c_N$
and coefficients $c_{N-1\pm 2q}\approx c_{N-1}$, belonging to two different
smooth (in index) sub--sequences; here, integer $q\sim {\cal O}\left(
N^0\right) $. Bearing this in mind, the large--$N$ version of recurrence
equation Eq. (\ref{q.09}) can be rewritten as 
\begin{equation}
\left[ \varepsilon ^2-\left( c_N^2+c_{N-1}^2\right) \right] P_N\left(
\varepsilon \right) =c_Nc_{N-1}\left[ P_{N-1}\left( \varepsilon \right)
+P_{N+1}\left( \varepsilon \right) \right] ,  \label{q.19}
\end{equation}
whence we get the following asymptotic identities: 
\begin{equation}
\varepsilon ^{2\lambda }P_N\left( \varepsilon \right) =\left(
c_N^2+c_{N-1}^2\right) ^\lambda \sum_{k=0}^\lambda \left( 
\begin{array}{c}
\lambda \\ 
k
\end{array}
\right) \left( \frac{c_Nc_{N-1}}{c_N^2+c_{N-1}^2}\right)
^k\sum_{j=0}^k\left( 
\begin{array}{c}
k \\ 
j
\end{array}
\right) P_{N+4j-2k}\left( \varepsilon \right) ,  \label{q.20}
\end{equation}
and 
\begin{equation}
\varepsilon ^{2\lambda +1}P_N\left( \varepsilon \right) =\left(
c_N^2+c_{N-1}^2\right) ^\lambda \sum_{k=0}^\lambda \left( 
\begin{array}{c}
\lambda \\ 
k
\end{array}
\right) \left( \frac{c_Nc_{N-1}}{c_N^2+c_{N-1}^2}\right)
^k\sum_{j=0}^k\left( 
\begin{array}{c}
k \\ 
j
\end{array}
\right) \left[ c_{N-1}P_{N+4j-2k+1}\left( \varepsilon \right)
+c_NP_{N+4j-2k-1}\left( \varepsilon \right) \right]  \label{q.21}
\end{equation}
with integer $\lambda \geq 0$.

Expansions Eqs. (\ref{q.20}) and (\ref{q.21}) make it possible to compute
the required functions ${\cal F}_N$ and ${\cal G}_N$ entering the
differential equation Eq. (\ref{q.13}) for fictitious wave functions in the
limit $N\gg 1$. Substituting the explicit form of the confinement potential
set by Eq. (\ref{q.02}) into Eqs. (\ref{q.11}) and (\ref{q.12}), we obtain 
\begin{equation}
A_N\left( \varepsilon \right) =2c_N\sum_{k=1}^pd_k\sum_{\lambda
=1}^{2k-1}\varepsilon ^{\lambda -1}\int d\mu \left( t\right) P_N^2\left(
t\right) t^{2k-\lambda -1},  \label{q.22}
\end{equation}
and 
\begin{equation}
B_N\left( \varepsilon \right) =2c_N\sum_{k=1}^pd_k\sum_{\lambda
=1}^{2k-1}\varepsilon ^{\lambda -1}\int d\mu \left( t\right) P_N\left(
t\right) P_{N-1}\left( t\right) t^{2k-\lambda -1},  \label{q.23}
\end{equation}
respectively. Both integrals above can be calculated using expansions Eqs. (%
\ref{q.20}), (\ref{q.21}), and exploiting the orthogonality expressed by Eq.
(\ref{q.08}). Detailed calculations, given in Appendices A and B, lead to
the following results: 
\begin{equation}
A_N\left( \varepsilon \right) =\frac 2\pi \left( {\cal D}_N^{+}-\left(
-1\right) ^N{\cal D}_N^{-}\right) {\cal P}\int_{{\cal D}_N^{-}}^{{\cal D}%
_N^{+}}\frac{dV}{dt}\frac{t^2}{t^2-\varepsilon ^2}\frac{dt}{\sqrt{\left[
\left( {\cal D}_N^{+}\right) ^2-t^2\right] \left[ t^2-\left( {\cal D}%
_N^{-}\right) ^2\right] }},  \label{q.24}
\end{equation}
\begin{equation}
B_N\left( \varepsilon \right) =\frac 2\pi \varepsilon {\cal P}\int_{{\cal D}%
_N^{-}}^{{\cal D}_N^{+}}\frac{dV}{dt}\frac{t^2-\left( -1\right) ^N{\cal D}%
_N^{-}{\cal D}_N^{+}}{\sqrt{\left[ \left( {\cal D}_N^{+}\right)
^2-t^2\right] \left[ t^2-\left( {\cal D}_N^{-}\right) ^2\right] }}\frac{dt}{%
t^2-\varepsilon ^2}-\frac{dV}{d\varepsilon }.  \label{q.25}
\end{equation}

Having obtained the explicit expressions for functions $A_N$ and $B_N$, it
is easy to verify that coefficients ${\cal F}_n\left( \varepsilon \right) $
and ${\cal G}_n\left( \varepsilon \right) $ entering the differential
equation Eq. (\ref{q.13}) for the fictitious wave function $\varphi
_n^{\left( \text{II}\right) }\left( \varepsilon \right) $ may be expressed
in terms of the Dyson density $\nu _D^{\left( \text{II}\right) }$ in the
two--cut phase supported on two disconnected intervals $\varepsilon \in
\left( -{\cal D}_N^{+},-{\cal D}_N^{-}\right) \cup \left( {\cal D}_N^{-},%
{\cal D}_N^{+}\right) $ 
\begin{equation}
\nu _D^{\left( \text{II}\right) }\left( \varepsilon \right) =\frac 2{\pi ^2}%
\left| \varepsilon \right| \sqrt{\left[ \left( {\cal D}_N^{+}\right)
^2-\varepsilon ^2\right] \left[ \varepsilon ^2-\left( {\cal D}_N^{-}\right)
^2\right] }{\cal P}\int_{{\cal D}_N^{-}}^{{\cal D}_N^{+}}dt\frac{dV/dt}{%
t^2-\varepsilon ^2}\frac 1{\sqrt{\left[ \left( {\cal D}_N^{+}\right)
^2-t^2\right] \left[ t^2-\left( {\cal D}_N^{-}\right) ^2\right] }}
\label{q.26}
\end{equation}
when $N\gg 1$. Namely, Eqs. (\ref{q.14}), (\ref{q.15}), (\ref{q.24}) and (%
\ref{q.25}) yield 
\begin{equation}
{\cal F}_N\left( \varepsilon \right) =\frac d{d\varepsilon }\log \left( 
\frac{\pi \left| \varepsilon \right| \nu _D^{\left( \text{II}\right) }\left(
\varepsilon \right) }{\sqrt{\left[ \left( {\cal D}_N^{+}\right)
^2-\varepsilon ^2\right] \left[ \varepsilon ^2-\left( {\cal D}_N^{-}\right)
^2\right] }}\right) ,  \label{q.27}
\end{equation}
\begin{equation}
{\cal G}_N\left( \varepsilon \right) =\left[ \pi \nu _D^{\left( \text{II}%
\right) }\left( \varepsilon \right) \right] ^2+\frac{\pi \nu _D^{\left( 
\text{II}\right) }\left( \varepsilon \right) }{\left| \varepsilon \right| 
\sqrt{\left[ \left( {\cal D}_N^{+}\right) ^2-\varepsilon ^2\right] \left[
\varepsilon ^2-\left( {\cal D}_N^{-}\right) ^2\right] }}\left[ \varepsilon
^2+\left( -1\right) ^N{\cal D}_N^{-}{\cal D}_N^{+}\right] .  \label{q.28}
\end{equation}
In the large--$N$ limit, the second term in Eq. (\ref{q.28}) can be
neglected provided $\varepsilon $ belongs to the one of allowed bands, so
that $\varphi _N^{\left( \text{II}\right) }\left( \varepsilon \right) $
satisfies the following asymptotic differential equation in the two--cut
phase: 
\begin{equation}
\frac{d^2\varphi _N^{\left( \text{II}\right) }\left( \varepsilon \right) }{%
d\varepsilon ^2}-\left[ \frac d{d\varepsilon }\log \left( \frac{\pi \left|
\varepsilon \right| \nu _D^{\left( \text{II}\right) }\left( \varepsilon
\right) }{\sqrt{\left[ \left( {\cal D}_N^{+}\right) ^2-\varepsilon ^2\right]
\left[ \varepsilon ^2-\left( {\cal D}_N^{-}\right) ^2\right] }}\right)
\right] \frac{d\varphi _N^{\left( \text{II}\right) }\left( \varepsilon
\right) }{d\varepsilon }+\left[ \pi \nu _D^{\left( \text{II}\right) }\left(
\varepsilon \right) \right] ^2\varphi _N^{\left( \text{II}\right) }\left(
\varepsilon \right) =0.  \label{q.29}
\end{equation}
We recall that ${\cal D}_N^{-}$ and ${\cal D}_N^{+}$ are the end points of
the eigenvalue support that obey the two integral equations 
\begin{equation}
\int_{{\cal D}_N^{-}}^{{\cal D}_N^{+}}\frac{dV}{dt}\frac{t^2dt}{\sqrt{\left[
\left( {\cal D}_N^{+}\right) ^2-t^2\right] \left[ t^2-\left( {\cal D}%
_N^{-}\right) ^2\right] }}=\frac{\pi N}2,  \label{ee.23}
\end{equation}
\begin{equation}
\int_{{\cal D}_N^{-}}^{{\cal D}_N^{+}}\frac{dV}{dt}\frac{dt}{\sqrt{\left[
\left( {\cal D}_N^{+}\right) ^2-t^2\right] \left[ t^2-\left( {\cal D}%
_N^{-}\right) ^2\right] }}=0,  \label{ee.24}
\end{equation}
obtained in Appendix C. One can verify that as ${\cal D}_N^{-}$ tends to
zero, we recover equation Eq. (\ref{q.16}) valid in the single--band regime.

\section{Local eigenvalue correlations}

Eigenvalue correlations in the spectra of two--band random matrices are
completely determined by the Dyson density of states entering the effective
Schr\"odinger equation Eq. (\ref{q.29}).

(i) In the spectrum bulk, the Dyson density is a well--behaved function that
can be taken approximately as a constant on the scale of a few eigenlevels.
Then, in the vicinity of some $\varepsilon _0$ that is chosen to be far
enough from the spectrum end points $\pm {\cal D}_N^{\pm }$, Eq. (\ref{q.29}%
) takes the form 
\begin{equation}
\frac{d^2\varphi _N^{\left( \text{II}\right) }\left( \varepsilon \right) }{%
d\varepsilon ^2}+\left[ \pi /\Delta \left( \varepsilon _0\right) \right]
^2\varphi _N^{\left( \text{II}\right) }\left( \varepsilon \right) =0,
\label{sin.1}
\end{equation}
with $\Delta \left( \varepsilon _0\right) =1/\nu _D^{\left( \text{II}\right)
}\left( \varepsilon _0\right) $ being the mean level spacing in the vicinity
of $\varepsilon _0$. Clearly, the universal sine law for the two--point
kernel, Eq. (\ref{q.06}), follows immediately.

(ii) Eigenvalue correlations near the end points of an eigenvalue support
are determined by the Dyson density as well. Noting that in the vicinity of $%
\left| \varepsilon \right| ={\cal D}_N^{\pm }$, the Dyson density can be
represented in the form \cite{BB,KF-1997}, 
\begin{equation}
\nu _D^{\left( \text{II}\right) }\left( \varepsilon \right) =\left[ \pm
\left( 1-\frac{\varepsilon ^2}{\left( {\cal D}_N^{\pm }\right) ^2}\right)
\right] ^{m+\frac 12}{\cal R}_N\left( \frac \varepsilon {{\cal D}_N^{\pm }}%
\right)  \label{soft.1}
\end{equation}
where ${\cal R}_N\left( \pm 1\right) \neq 0$ and $m$ is the order of
multicriticality, we readily recover the universal multicritical
correlations previously found \cite{KF-1997} in the soft--edge scaling limit
for $%
\mathop{\rm U}
\left( N\right) $ invariant matrix model in the single--band phase.

\section{Smoothed connected `density--density' correlator}

Let us turn to the study of the connected `density--density' correlator that
is expressed in terms of the two--point kernel, Eq. (\ref{q.06}), as follows 
\begin{equation}
\left\langle \delta \nu _N\left( \varepsilon \right) \delta \nu _N\left(
\varepsilon ^{\prime }\right) \right\rangle _{\text{II}}=-\frac{c_N^2}{%
\left( \varepsilon -\varepsilon ^{\prime }\right) ^2}\left\{ \varphi
_N^2\left( \varepsilon \right) \varphi _{N-1}^2\left( \varepsilon ^{\prime
}\right) +\varphi _N^2\left( \varepsilon ^{\prime }\right) \varphi
_{N-1}^2\left( \varepsilon \right) -2\varphi _N\left( \varepsilon \right)
\varphi _{N-1}\left( \varepsilon \right) \varphi _N\left( \varepsilon
^{\prime }\right) \varphi _{N-1}\left( \varepsilon ^{\prime }\right) \right\}
\label{ee.33}
\end{equation}
where $\varepsilon \neq \varepsilon ^{\prime }$, and the upper index $\left( 
\text{II}\right) $ in $\varphi _n$ has been omitted for brevity. We still
deal with the two--band phase. Expression Eq. (\ref{ee.33}) contains rapid
oscillations on the scale of the mean level spacing. These oscillations are
due to presence in Eq. (\ref{ee.33}) of oscillating wave functions $\varphi
_N$ and $\varphi _{N-1}$.

To average over the rapid oscillations, we integrate, over the entire real
axis, rapidly varying wave functions in Eq. (\ref{ee.33}) multiplied by an
arbitrary, smooth, slowly varying function. To illustrate the idea, consider
the integral 
\begin{equation}
I_{{\sf f}}=\int_{-\infty }^{+\infty }d\varepsilon \varphi _N^2\left(
\varepsilon \right) {\sf f}\left( \varepsilon \right) ,  \label{ee.34}
\end{equation}
where ${\sf f}\left( \varepsilon \right) $ is arbitrary slowly varying
function that should be chosen to be even due to the evenness of $\varphi
_N^2\left( \varepsilon \right) $. Setting 
\begin{equation}
{\sf f}\left( \varepsilon \right) =\sum_{\alpha =0}^\infty {\sf f}_\alpha
\varepsilon ^{2\alpha },  \label{ee.35}
\end{equation}
we immediately obtain with the help of Eqs. (A1) and (A6) that 
\begin{equation}
I_{{\sf f}}=\sum_{\alpha =0}^\infty {\sf f}_\alpha \Lambda _{2\alpha }=\frac 
2\pi \int_{{\cal D}_N^{-}}^{{\cal D}_N^{+}}\frac{\varepsilon {\sf f}\left(
\varepsilon \right) d\varepsilon }{\sqrt{\left[ \left( {\cal D}_N^{+}\right)
^2-\varepsilon ^2\right] \left[ \varepsilon ^2-\left( {\cal D}_N^{-}\right)
^2\right] }}.  \label{ee.37}
\end{equation}
Bearing in mind that both ${\sf f}\left( \varepsilon \right) $ and $\varphi
_N^2\left( \varepsilon \right) $ are even functions, the last integral can
be transformed as follows 
\begin{equation}
\ I_{{\sf f}}=\int_{-\infty }^{+\infty }d\varepsilon \varphi _N^2\left(
\varepsilon \right) {\sf f}\left( \varepsilon \right) =\frac 1\pi \int_{%
{\cal D}_N^{-}<\left| \varepsilon \right| <{\cal D}_N^{+}}\frac{\left|
\varepsilon \right| {\sf f}\left( \varepsilon \right) d\varepsilon }{\sqrt{%
\left[ \left( {\cal D}_N^{+}\right) ^2-\varepsilon ^2\right] \left[
\varepsilon ^2-\left( {\cal D}_N^{-}\right) ^2\right] }},  \label{ee.38}
\end{equation}
whence we conclude that in the large--$N$ limit, 
\begin{equation}
\overline{\varphi _N^2\left( \varepsilon \right) }=\frac 1\pi \frac{\left|
\varepsilon \right| }{\sqrt{\left[ \left( {\cal D}_N^{+}\right)
^2-\varepsilon ^2\right] \left[ \varepsilon ^2-\left( {\cal D}_N^{-}\right)
^2\right] }}\Theta \left( {\cal D}_N^{+}-\left| \varepsilon \right| \right)
\Theta \left( \left| \varepsilon \right| -{\cal D}_N^{-}\right) .
\label{ee.39}
\end{equation}

The same procedure should be carried out with expression $\varphi _N\left(
\varepsilon \right) \varphi _{N-1}\left( \varepsilon \right) $ in Eq. (\ref
{ee.33}). Since this construction is an odd function of $\varepsilon $, we
have to consider the integral 
\begin{equation}
I_{{\sf g}}=\int_{-\infty }^{+\infty }d\varepsilon \varphi _N\left(
\varepsilon \right) \varphi _{N-1}\left( \varepsilon \right) {\sf g}\left(
\varepsilon \right) ,  \label{ee.40}
\end{equation}
with 
\begin{equation}
{\sf g}\left( \varepsilon \right) =\sum_{\alpha =0}^\infty {\sf g}_\alpha
\varepsilon ^{2\alpha +1}  \label{ee.41}
\end{equation}
being a smooth odd function. It is easy to see with the help of Eqs. (B1),
(B5) and (C7) that 
\begin{equation}
I_{{\sf g}}=\sum_{\alpha =0}^\infty {\sf g}_\alpha \Gamma _{2\alpha +1}=%
\frac 2{\pi \left[ {\cal D}_N^{+}-\left( -1\right) ^N{\cal D}_N^{-}\right] }%
\int_{{\cal D}_N^{-}}^{{\cal D}_N^{+}}\frac{{\sf g}\left( \varepsilon
\right) \left[ \varepsilon ^2-\left( -1\right) ^N{\cal D}_N^{-}{\cal D}%
_N^{+}\right] d\varepsilon }{\sqrt{\left[ \left( {\cal D}_N^{+}\right)
^2-\varepsilon ^2\right] \left[ \varepsilon ^2-\left( {\cal D}_N^{-}\right)
^2\right] }}.  \label{ee.43}
\end{equation}
Exploiting the oddness of ${\sf g}\left( \varepsilon \right) $ and $\varphi
_N\left( \varepsilon \right) \varphi _{N-1}\left( \varepsilon \right) $, we
write Eq. (\ref{ee.43}) in the form 
\begin{eqnarray}
I_{{\sf g}} &=&\int_{-\infty }^{+\infty }d\varepsilon \varphi _N\left(
\varepsilon \right) \varphi _{N-1}\left( \varepsilon \right) {\sf g}\left(
\varepsilon \right) =\frac 1{\pi \left[ {\cal D}_N^{+}-\left( -1\right) ^N%
{\cal D}_N^{-}\right] }  \nonumber \\
&&\times \int_{{\cal D}_N^{-}<\left| \varepsilon \right| <{\cal D}_N^{+}}%
\frac{{\sf g}\left( \varepsilon \right) \left[ \varepsilon ^2-\left(
-1\right) ^N{\cal D}_N^{-}{\cal D}_N^{+}\right] 
\mathop{\rm sgn}
\left( \varepsilon \right) d\varepsilon }{\sqrt{\left[ \left( {\cal D}%
_N^{+}\right) ^2-\varepsilon ^2\right] \left[ \varepsilon ^2-\left( {\cal D}%
_N^{-}\right) ^2\right] }}.  \label{ee.44}
\end{eqnarray}
Equation (\ref{ee.44}) leads us to the conclusion that in the large--$N$
limit, 
\begin{equation}
\overline{\varphi _N\left( \varepsilon \right) \varphi _{N-1}\left(
\varepsilon \right) }=\frac{%
\mathop{\rm sgn}
\left( \varepsilon \right) }{\pi \left[ {\cal D}_N^{+}-\left( -1\right) ^N%
{\cal D}_N^{-}\right] }\frac{\varepsilon ^2-\left( -1\right) ^N{\cal D}_N^{-}%
{\cal D}_N^{+}}{\sqrt{\left[ \left( {\cal D}_N^{+}\right) ^2-\varepsilon
^2\right] \left[ \varepsilon ^2-\left( {\cal D}_N^{-}\right) ^2\right] }}%
\Theta \left( {\cal D}_N^{+}-\left| \varepsilon \right| \right) \Theta
\left( \left| \varepsilon \right| -{\cal D}_N^{-}\right) .  \label{ee.45}
\end{equation}

Combining Eqs. (\ref{ee.33}), (\ref{ee.39}), (\ref{ee.45}) and (C7), we
finally arrive at the following formula for smoothed `density--density'
correlator, 
\begin{eqnarray}
\overline{\left\langle \delta \nu _N\left( \varepsilon \right) \delta \nu
_N\left( \varepsilon ^{\prime }\right) \right\rangle }_{\text{II}} &=&-\frac{%
\mathop{\rm sgn}
\left( \varepsilon \varepsilon ^{\prime }\right) }{2\pi ^2}\Theta \left( 
{\cal D}_N^{+}-\left| \varepsilon \right| \right) \Theta \left( \left|
\varepsilon \right| -{\cal D}_N^{-}\right) \Theta \left( {\cal D}%
_N^{+}-\left| \varepsilon ^{\prime }\right| \right) \Theta \left( \left|
\varepsilon ^{\prime }\right| -{\cal D}_N^{-}\right)  \label{y1} \\
&&\times \left\{ \frac 1{\left( \varepsilon -\varepsilon ^{\prime }\right) ^2%
}\frac{\left[ \varepsilon \varepsilon ^{\prime }-\left( {\cal D}%
_N^{-}\right) ^2\right] \left[ \left( {\cal D}_N^{+}\right) ^2-\varepsilon
\varepsilon ^{\prime }\right] }{\sqrt{\left[ \left( {\cal D}_N^{+}\right)
^2-\varepsilon ^2\right] \left[ \varepsilon ^2-\left( {\cal D}_N^{-}\right)
^2\right] }\sqrt{\left[ \left( {\cal D}_N^{+}\right) ^2-\varepsilon ^{\prime
2}\right] \left[ \varepsilon ^{\prime 2}-\left( {\cal D}_N^{-}\right)
^2\right] }}\right.  \nonumber \\
&&+\left. \left( -1\right) ^N\frac{{\cal D}_N^{-}{\cal D}_N^{+}}{\sqrt{%
\left[ \left( {\cal D}_N^{+}\right) ^2-\varepsilon ^2\right] \left[
\varepsilon ^2-\left( {\cal D}_N^{-}\right) ^2\right] }\sqrt{\left[ \left( 
{\cal D}_N^{+}\right) ^2-\varepsilon ^{\prime 2}\right] \left[ \varepsilon
^{\prime 2}-\left( {\cal D}_N^{-}\right) ^2\right] }}\right\} .  \nonumber
\label{y2}
\end{eqnarray}
The same formula can be obtained by WKB by solving Eq. (\ref{q.29}), using
definition Eq. (\ref{ee.33}) followed by averaging over rapid oscillations.
It can be verified that for $N$ even, this result coincides with Eq. (6.6)
of Ref. \cite{Deo-1997} where it was obtained by a completely different
method, and for the case of odd $N$ being omitted.

It is seen from Eq. (\ref{y1}) that smoothed `density--density' correlator
in the two--band phase is a new universal function in random matrix theory.
It is universal in the sense that the information of the distribution Eq. (%
\ref{q.01}) is encoded into the `density--density' correlator only through
the end points ${\cal D}_N^{\pm }$ of the eigenvalue support. A striking
feature of the new universal function Eq. (\ref{y1}) is its {\it sharp}
dependence on the oddness/evenness of the dimension $N$ of the random
matrices whose spectra are {\it bounded}. The origin of this unusual large--$%
N$ behavior will be discussed in the next Section.

Finally, let us speculate about the universal correlator Eq. (\ref{y1}) in
the limit of {\it unbounded} spectrum, ${\cal D}_N^{+}\rightarrow \infty $,
with a gap. Inasmuch as it describes correlations between the eigenlevels
which are repelled from each other in accordance with the logarithmic law,
that is known to be realized \cite{AS-1986,JPB-1993} in the weakly
disordered systems on the energy scale $\left| \varepsilon -\varepsilon
^{\prime }\right| \ll E_c$ ($E_c$ is the Thouless energy), we may {\it %
conjecture }that the corresponding limiting universal expression 
\begin{equation}
\lim_{{\cal D}_N^{+}\rightarrow +\infty }\overline{\left\langle \delta \nu
_N\left( \varepsilon \right) \delta \nu _N\left( \varepsilon ^{\prime
}\right) \right\rangle }_{\text{II}}=-\frac{%
\mathop{\rm sgn}
\left( \varepsilon \varepsilon ^{\prime }\right) }{2\pi ^2\left( \varepsilon
-\varepsilon ^{\prime }\right) ^2}\frac{\varepsilon \varepsilon ^{\prime
}-\Delta ^2}{\sqrt{\left[ \varepsilon ^2-\Delta ^2\right] \left[ \varepsilon
^{\prime 2}-\Delta ^2\right] }}\Theta \left( \left| \varepsilon \right|
-\Delta \right) \Theta \left( \left| \varepsilon ^{\prime }\right| -\Delta
\right) ,  \label{ee.48}
\end{equation}
reflects the universal properties of {\it real} chaotic systems with a
forbidden gap $\Delta ={\cal D}_N^{-}$ and broken time reversal symmetry,
provided $\left| \varepsilon -\varepsilon ^{\prime }\right| \ll E_c$. In two
limiting situations (i) of gapless spectrum, $\Delta =0$, and (ii) far from
the gap, $\left| \varepsilon \right| ,\left| \varepsilon ^{\prime }\right|
\gg \Delta $, the correlator Eq. (\ref{ee.48}) coincides with that known in
the random matrix theory of gapless ensembles \cite
{BrezinZee1993,Beenakker1994} and derived in Ref. \cite{AS-1986} within the
framework of diagrammatic technique for spectrum of electron in a random
impurity potential.

\section{Concluding remarks}

In this study we developed a unified formalism allowing the computation of
both global and local spectral characteristics of $%
\mathop{\rm U}
\left( N\right) $ invariant ensembles of large random matrices possessing $%
Z2 $--symmetry, and deformed in such a way that their spectra contain a
forbidden gap. We proved that in the pure two--band phase, the local
eigenvalue correlations are insensitive to this deformation both in the bulk
and soft--edge scaling limits. In contrast, global smoothed eigenvalue
correlations in the two--band phase differ drastically from those in the
single--band phase, and generically satisfy a new universal law, Eq. (\ref
{y1}), which is unusually sensitive to the oddness/evenness of the random
matrix dimension if the spectrum support is {\it bounded}. On the formal
level, this sensitivity is a direct consequence of the `period--two'
behavior \cite{Molinari1988,Demeterfi-1990} of the recurrence coefficients $%
c_n$ [see Eq. (\ref{q.09})] that is characteristic of two--band phase of
reduced Hermitian matrix model. To see this, consider the simplest connected
correlator $\left\langle 
\mathop{\rm Tr}
{\bf H}%
\mathop{\rm Tr}
{\bf H}\right\rangle _c$ that can be {\it exactly} represented in terms of
recurrence coefficients for any $n$, 
\begin{equation}
\left\langle 
\mathop{\rm Tr}
{\bf H}%
\mathop{\rm Tr}
{\bf H}\right\rangle _c=c_n^2\text{.}  \label{e.100}
\end{equation}
Since in the two--band phase $c_n$ is a double--valued function of index $n$%
, alternating between two different functions as $n$ goes from odd to even,
the large--$N$ limit of the correlator $\left\langle 
\mathop{\rm Tr}
{\bf H}%
\mathop{\rm Tr}
{\bf H}\right\rangle _c$ strongly depends on whether infinity is approached
through odd or even $N$. Then, an implementation of a double--valued
behavior of $c_n$ into the higher order correlators of the form $%
\left\langle 
\mathop{\rm Tr}
{\bf H}^k%
\mathop{\rm Tr}
{\bf H}^l\right\rangle _c$ contributing to the connected `density--density'
correlator gives rise to the new universal expression Eq. (\ref{y1}).

Let us, however, point out that no such sensitivity has been detected in a
number of previous studies \cite{AmbjornAkemann1996,Akemann1996} exploiting
a loop--equation technique. This is due to the following reasons. In the
method of loop equations, used for a treatment of non--Gaussian random
matrix ensembles fallen in a multi--band phase, one is forced to keep the
most general (non--symmetric) confinement potential $V\left( \varepsilon
\right) =\sum_{k=1}^{2p}\widetilde{d}_k\varepsilon ^k/k$ until very end of
the calculations, leading to a necessity to take the thermodynamic limit $%
N\rightarrow \infty $ prior to any others. Therefore, $Z2$--symmetry in this
calculational scheme can only be implemented by restoring $Z2$--symmetry at
the final stage of the calculations, setting all the extra coupling
constants $\widetilde{d}_{2k+1}$ to zero. Doing so, one arrives at the
results reported in Refs. \cite{AmbjornAkemann1996,Akemann1996}.

>From this point of view, the formalism developed in this paper corresponds
to the {\it opposite sequence }of thermodynamic and $Z2$--symmetry limits,
since we have considered the random matrix model that possesses $Z2$%
--symmetry from the beginning. Qualitatively different large--$N$ behavior
of the smoothed connected `density--density' correlator, Eq. (\ref{y1}), and
of the smoothed connected two--point Green's function given by Eq. (15) of
Ref. \cite{Akemann1996} provides a direct evidence that the order of
thermodynamic and $Z2$--symmetry limits is indeed important when studying
global spectral characteristics of multi--band random matrices.

\vspace{0.8cm}

\begin{center}
{\bf Acknowledgments}
\end{center}

The authors thank G. Akemann and N. Deo for useful discussions. Financial
support (EK) from the Ministry of Science of Israel through the Levy Eshkol
Fellowship is gratefully acknowledged.

\newpage\ 

\begin{center}
{\bf Appendix A: Calculation of the function }$A_N\left( \varepsilon \right) 
$
\end{center}

Let us consider an integral 
\begin{equation}
\Lambda _{2\sigma }=\int d\mu \left( t\right) P_N^2\left( t\right)
t^{2\sigma }  \eqnum{A1}
\end{equation}
with integer $\sigma \geq 0$. Making use of Eq. (\ref{q.20}), we rewrite $%
\Lambda _{2\sigma }$ in the form 
\begin{equation}
\Lambda _{2\sigma }=\left( c_N^2+c_{N-1}^2\right) ^\sigma \sum_{k=0}^\sigma
\left( 
\begin{array}{c}
\sigma \\ 
k
\end{array}
\right) \left( \frac{c_Nc_{N-1}}{c_N^2+c_{N-1}^2}\right)
^k\sum_{j=0}^k\left( 
\begin{array}{c}
k \\ 
j
\end{array}
\right) \int d\mu \left( t\right) P_N\left( t\right) P_{N+4j-2k}\left(
t\right) .  \eqnum{A2}
\end{equation}
Orthogonality of the $P_n$ allows us to integrate over the measure $d\mu $,
thus simplifying Eq. (A2): 
\begin{equation}
\Lambda _{2\sigma }=\left( c_N^2+c_{N-1}^2\right) ^\sigma \sum_{k=0}^\sigma
\left( 
\begin{array}{c}
\sigma \\ 
k
\end{array}
\right) \left( \frac{c_Nc_{N-1}}{c_N^2+c_{N-1}^2}\right)
^k\sum_{j=0}^k\left( 
\begin{array}{c}
k \\ 
j
\end{array}
\right) \delta _{2j}^k,  \eqnum{A3}
\end{equation}
where $\delta _{k^{\prime }}^k$ is the Kronecker symbol. Using integral
representation 
\begin{equation}
\delta _{k^{\prime }}^k=%
\mathop{\rm Re}
\int_0^{2\pi }\frac{d\theta }{2\pi }\exp \left\{ i\left( k-k^{\prime
}\right) \theta \right\} \text{,}  \eqnum{A4}
\end{equation}
one can perform the double summation in Eq. (A3): 
\begin{equation}
\Lambda _{2\sigma }=\int_0^{2\pi }\frac{d\theta }{2\pi }\left(
c_N^2+c_{N-1}^2+2c_Nc_{N-1}\cos \theta \right) ^\sigma \text{.}  \eqnum{A5}
\end{equation}
Introducing a new integration variable $t^2=c_N^2+c_{N-1}^2+2c_Nc_{N-1}\cos
\theta $, we derive an integral formula 
\begin{equation}
\Lambda _{2\sigma }=\frac 2\pi \int_{{\cal D}_N^{-}}^{{\cal D}_N^{+}}\frac{%
t^{2\sigma +1}dt}{\sqrt{\left[ \left( {\cal D}_N^{+}\right) ^2-t^2\right]
\left[ t^2-\left( {\cal D}_N^{-}\right) ^2\right] }}  \eqnum{A6}
\end{equation}
with 
\begin{equation}
{\cal D}_N^{\pm }=\left| c_N\pm c_{N-1}\right| .  \eqnum{A7}
\end{equation}

Now, taking into account representation Eq. (A6) for $\Lambda _{2\sigma }$,
and using the fact that $\Lambda _{2\sigma +1}\equiv 0$, we obtain from Eq. (%
\ref{q.22}) 
\begin{equation}
A_N\left( \varepsilon \right) =2c_N\sum_{k=1}^pd_k\sum_{\sigma =1}^k\Lambda
_{2\left( k-\sigma \right) }\varepsilon ^{2\sigma -2}.  \eqnum{A8}
\end{equation}
Summing over $\sigma $ yields 
\begin{equation}
A_N\left( \varepsilon \right) =\frac{4c_N}\pi \sum_{k=1}^pd_k\int_{{\cal D}%
_N^{-}}^{{\cal D}_N^{+}}\frac{tdt}{\sqrt{\left[ \left( {\cal D}_N^{+}\right)
^2-t^2\right] \left[ t^2-\left( {\cal D}_N^{-}\right) ^2\right] }}\frac{%
t^{2k}-\varepsilon ^{2k}}{t^2-\varepsilon ^2},  \eqnum{A9}
\end{equation}
whence we get, with the help of Eq. (\ref{q.02}), 
\begin{equation}
A_N\left( \varepsilon \right) =\frac{4c_N}\pi \int_{{\cal D}_N^{-}}^{{\cal D}%
_N^{+}}\frac{dt}{\sqrt{\left[ \left( {\cal D}_N^{+}\right) ^2-t^2\right]
\left[ t^2-\left( {\cal D}_N^{-}\right) ^2\right] }}\frac t{t^2-\varepsilon
^2}\left( t\frac{dV}{dt}-\varepsilon \frac{dV}{d\varepsilon }\right) . 
\eqnum{A10}
\end{equation}
Further, noting that 
\begin{equation}
{\cal P}\int_{{\cal D}_N^{-}}^{{\cal D}_N^{+}}\frac{dt}{\sqrt{\left[ \left( 
{\cal D}_N^{+}\right) ^2-t^2\right] \left[ t^2-\left( {\cal D}_N^{-}\right)
^2\right] }}\frac t{t^2-\varepsilon ^2}\equiv 0,  \eqnum{A11}
\end{equation}
and taking into account Eq. (C7), leads to the final expression given by Eq.
(\ref{q.24}).\vspace{0.2in}

\begin{center}
{\bf Appendix B:\ Calculation\ of\ the\ function }$B_N\left( \varepsilon
\right) $
\end{center}

Let us consider an integral 
\begin{equation}
\Gamma _{2\sigma +1}=\int d\mu \left( t\right) P_N\left( t\right)
P_{N-1}\left( t\right) t^{2\sigma +1}  \eqnum{B1}
\end{equation}
with integer $\sigma \geq 0$. Making use of expansion Eq. (\ref{q.21}), we
rewrite Eq. (B1) in the form that allows us to perform the integration over
the measure $d\mu $: 
\begin{eqnarray}
\Gamma _{2\sigma +1} &=&\frac 12\left( c_N^2+c_{N-1}^2\right) ^\sigma \int
d\mu \left( t\right) P_{N-1}\left( t\right) \sum_{k=0}^\sigma \left( 
\begin{array}{c}
\sigma \\ 
k
\end{array}
\right) \left( \frac{c_Nc_{N-1}}{c_N^2+c_{N-1}^2}\right)
^k\sum_{j=0}^k\left( 
\begin{array}{c}
k \\ 
j
\end{array}
\right)  \nonumber \\
&&\ \times \left[ c_{N-1}P_{N+4j-2k+1}\left( t\right)
+c_NP_{N+4j-2k-1}\left( t\right) \right] .  \eqnum{B2}
\end{eqnarray}
After integration, we get 
\begin{equation}
\Gamma _{2\sigma +1}=\frac 12\left( c_N^2+c_{N-1}^2\right) ^\sigma
\sum_{k=0}^\sigma \left( 
\begin{array}{c}
\sigma \\ 
k
\end{array}
\right) \left( \frac{c_Nc_{N-1}}{c_N^2+c_{N-1}^2}\right)
^k\sum_{j=0}^k\left( 
\begin{array}{c}
k \\ 
j
\end{array}
\right) \left[ c_{N-1}\delta _{2j+1}^k+c_N\delta _{2j}^k\right] .  \eqnum{B3}
\end{equation}
The double summation in Eq. (B3) can be performed using the integral
representation for the Kronecker symbol given by Eq. (A4): 
\begin{equation}
\Gamma _{2\sigma +1}=\frac 12\int_0^{2\pi }\frac{d\theta }{2\pi }\left(
c_N^2+c_{N-1}^2+2c_Nc_{N-1}\cos \theta \right) ^\sigma \left[
c_N+c_{N-1}\cos \theta \right] .  \eqnum{B4}
\end{equation}
Introducing a new integration variable $t^2=c_N^2+c_{N-1}^2+2c_Nc_{N-1}\cos
\theta $, we get 
\begin{equation}
\Gamma _{2\sigma +1}=\frac 1{\pi c_N}\int_{{\cal D}_N^{-}}^{{\cal D}_N^{+}}%
\frac{t^{2\sigma +1}dt}{\sqrt{\left[ \left( {\cal D}_N^{+}\right)
^2-t^2\right] \left[ t^2-\left( {\cal D}_N^{-}\right) ^2\right] }}\left[
t^2+c_N^2-c_{N-1}^2\right] .  \eqnum{B5}
\end{equation}
Then, Eqs. (\ref{q.23}), (B1) and (B5) yield 
\begin{equation}
B_N\left( \varepsilon \right) =\frac 2\pi \sum_{k=1}^pd_k\sum_{\sigma
=1}^{k-1}\Gamma _{2k-2\sigma -1}\varepsilon ^{2\sigma -1}.  \eqnum{B6}
\end{equation}
Summing over $\sigma $ leads to the integral expression 
\begin{eqnarray}
B_N\left( \varepsilon \right) &=&\frac 2\pi \sum_{k=1}^pd_k\int_{{\cal D}%
_N^{-}}^{{\cal D}_N^{+}}dt\frac{\left[ t^2+c_N^2-c_{N-1}^2\right] }{\sqrt{%
\left[ \left( {\cal D}_N^{+}\right) ^2-t^2\right] \left[ t^2-\left( {\cal D}%
_N^{-}\right) ^2\right] }}\frac{\varepsilon t^{2k-1}-t\varepsilon ^{2k-1}}{%
t^2-\varepsilon ^2}  \nonumber \\
&=&\frac 2\pi \int_{{\cal D}_N^{-}}^{{\cal D}_N^{+}}\frac{dt}{\sqrt{\left[
\left( {\cal D}_N^{+}\right) ^2-t^2\right] \left[ t^2-\left( {\cal D}%
_N^{-}\right) ^2\right] }}\frac{t^2+c_N^2-c_{N-1}^2}{t^2-\varepsilon ^2}%
\left( \varepsilon \frac{dV}{dt}-t\frac{dV}{d\varepsilon }\right) . 
\eqnum{B7}
\end{eqnarray}
Now, taking into account Eqs. (A11) and (C6), we obtain Eq. (\ref{q.25}).%
\vspace{0.2in}

\begin{center}
{\bf Appendix C:\ Soft edges of eigenvalue support}
\end{center}

To find the equations determining the points ${\cal D}_N^{\pm }$ where the
Dyson spectral density goes to zero, we start with the following formula
from the theory of orthogonal polynomials \cite{Nevai-1986} 
\begin{equation}
n=2c_n\int d\mu \left( t\right) \frac{dV}{dt}P_n\left( t\right)
P_{n-1}\left( t\right) .  \eqnum{C1}
\end{equation}
Let us use expansion Eq. (\ref{q.21}) to calculate asymptotically the
integral entering Eq. (C1) in the limit $n=N\gg 1$. It is easy to see that 
\begin{equation}
N=2c_N\sum_{\lambda =1}^pd_\lambda \int d\mu \left( t\right) P_N\left(
t\right) P_{N-1}\left( t\right) t^{2\lambda -1}=2c_N\sum_{\lambda
=1}^pd_\lambda \Gamma _{2\lambda -1},  \eqnum{C2}
\end{equation}
where $\Gamma _{2\lambda -1}$ is given by Eq. (B5). Then, we immediately
obtain the relationship 
\begin{equation}
N=\frac 2\pi \int_{{\cal D}_N^{-}}^{{\cal D}_N^{+}}\frac{dt}{\sqrt{\left[
\left( {\cal D}_N^{+}\right) ^2-t^2\right] \left[ t^2-\left( {\cal D}%
_N^{-}\right) ^2\right] }}\frac{dV}{dt}\left[ t^2+c_N^2-c_{N-1}^2\right] . 
\eqnum{C3}
\end{equation}
This result rewritten for $n=N-1$, yields in the large--$N$ limit, 
\begin{equation}
N=\frac 2\pi \int_{{\cal D}_N^{-}}^{{\cal D}_N^{+}}\frac{dt}{\sqrt{\left[
\left( {\cal D}_N^{+}\right) ^2-t^2\right] \left[ t^2-\left( {\cal D}%
_N^{-}\right) ^2\right] }}\frac{dV}{dt}\left[ t^2+c_{N-1}^2-c_N^2\right] . 
\eqnum{C4}
\end{equation}
Equations (C3) and (C4) yield two equations whose solutions determine the
edge points ${\cal D}_N^{\pm }$: 
\begin{equation}
\int_{{\cal D}_N^{-}}^{{\cal D}_N^{+}}\frac{t^2dt}{\sqrt{\left[ \left( {\cal %
D}_N^{+}\right) ^2-t^2\right] \left[ t^2-\left( {\cal D}_N^{-}\right)
^2\right] }}\frac{dV}{dt}=\frac{\pi N}2,  \eqnum{C5}
\end{equation}
and 
\begin{equation}
\int_{{\cal D}_N^{-}}^{{\cal D}_N^{+}}\frac{dt}{\sqrt{\left[ \left( {\cal D}%
_N^{+}\right) ^2-t^2\right] \left[ t^2-\left( {\cal D}_N^{-}\right)
^2\right] }}\frac{dV}{dt}=0.  \eqnum{C6}
\end{equation}

\narrowtext
\noindent 
Finally, we note that because $P_{-1}\left( \varepsilon \right) =0$, it
follows from Eq. (\ref{q.09}) that $c_0=0$, and as a consequence, an even
branch $c_{2N}$ always lies lower than an odd branch $c_{2N\pm 1}$, so that $%
c_{2N}<c_{2N\pm 1}$. Then, we may conclude from Eq. (A7) that 
\begin{equation}
c_N=\frac{{\cal D}_N^{+}-\left( -1\right) ^N{\cal D}_N^{-}}2.  \eqnum{C7}
\end{equation}

\widetext

\end{document}